\documentclass[twocolumn]{aastex62}

\definecolor{blue-violet}{rgb}{0.54, 0.17, 0.89}

\usepackage{multirow}
\usepackage{makecell}

\graphicspath{{./}{figures/}}

\shorttitle{Tidally Induced Features in Cluster Dwarfs}
\shortauthors{Smith et al.}

\begin{document}

\title{Brought to Light II: Revealing the Origins of Cloaked Spiral Features in Cluster Passive Dwarf Galaxies}

\correspondingauthor{Rory Smith}
\email{rory.smith@kasi.re.kr}

\author{Rory Smith}
\affil{Korea Astronomy and Space Science Institute (KASI), 776 Daedeokdae-ro, Yuseong-gu, Daejeon 34055, Korea}

\author{Josefina Michea}
\affiliation{Astronomisches Rechen-Institut, Zentrum f\"ur Astronomie der Universit\"at Heidelberg, M\"onchhofstra{\ss}e 12-14, 69120 Heidelberg, Germany}

\author{Anna Pasquali}
\affiliation{Astronomisches Rechen-Institut, Zentrum f\"ur Astronomie der Universit\"at Heidelberg, M\"onchhofstra{\ss}e 12-14, 69120 Heidelberg, Germany}

\author{Paula Calder\'on-Castillo}
\affiliation{Departamento de Astronom\'ia, Universidad de Concepci\'on, Casilla 160-C, Concepci\'on, Chile}

\author{Katarina Kraljic}
\affiliation{Aix Marseille Universit\'e, CNRS, CNES, UMR 7326, Laboratoire d'Astrophysique de Marseille, Marseille, France}

\author{Sanjaya Paudel}
\affiliation{Yonsei University, Graduate School of Earth System Sciences-Astronomy-Atmospheric Sciences, Yonsei-ro 50, Seoul 120-749, Korea}

\author{Thorsten Lisker}
\affiliation{Astronomisches Rechen-Institut, Zentrum f\"ur Astronomie der Universit\"at Heidelberg, M\"onchhofstra{\ss}e 12-14, 69120 Heidelberg, Germany}

\author{Jihye Shin}
\affiliation{Korea Astronomy and Space Science Institute (KASI), 776 Daedeokdae-ro, Yuseong-gu, Daejeon 34055, Korea}

\author{Jongwan Ko}
\affiliation{Korea Astronomy and Space Science Institute (KASI), 776 Daedeokdae-ro, Yuseong-gu, Daejeon 34055, Korea}
\affiliation{University of Science and Technology (UST), Daejeon 34113, Korea}

\author{Reynier Peletier}
\affiliation{Kapteyn Astronomical Institute, University of Groningen, Landleven 12, NL-9747 AD Groningen, The Netherlands}

\author{Eva K. Grebel}
\affiliation{Astronomisches Rechen-Institut, Zentrum f\"ur Astronomie der Universit\"at Heidelberg, M\"onchhofstra{\ss}e 12-14, 69120 Heidelberg, Germany}



\begin{abstract}

In our companion paper (Brought to Light I: Michea et al. 2021), we reveal spectacular spiral galaxy-like features in deep optical imaging of nine Virgo early-type dwarf galaxies, hidden beneath a dominating smooth stellar disk. Using a new combination of approaches, we find that bar- and spiral-like features contribute 2.2-6.4\% of the total flux within 2 R$_{\rm{eff}}$. In this study, we conduct high resolution simulations of cluster harassment of passive dwarf galaxies. Following close pericenter passages of the cluster core, tidal triggering generates features in our model disks that bear a striking resemblance to the observed features. However, we find the disks must be highly rotationally supported (V$_{\rm{peak}}/\sigma_0 \sim 3$), much higher than typically observed. We propose that some early-type dwarfs may contain a few percent of their mass in a cold, thin disk which is buried in the light of a hot, diffuse disk, and only revealed when they undergo tidal triggering. The red optical colors of our sample do not indicate any recent significant star formation, and our simulations show that very plunging pericenter passages (r$_{\rm{peri}}<0.25$r$_{\rm{vir}}$) are required for tidal triggering. Thus, many cluster early-type dwarfs with less plunging orbits may host a yet undetected cold stellar disk component. We discuss possible origin scenarios and consider why similar mass star-forming galaxies in the field are significantly more thin disk dominated than in our cluster sample.
\end{abstract}


\section{Introduction} \label{sec:intro}

By number, dwarf early-type galaxies (ETGs) dominate the galaxy population in rich clusters \citep{VCC}. They are characterized as quiescent galaxies, with low stellar masses ($M_{\rm{str}}<$10$^{9-9.5}$~M$_\odot$), and lacking in significant quantities of gas and dust. Their light profiles are generally quite smooth with roughly exponential light profiles. Given their low masses and thus shallow potential wells, they are expected to be more sensitive to environmental effects than giant galaxies.

However, recent studies suggest they are not as simple as once thought. Most cluster dwarf ETGs are not well fit by a single S\'ersic profile, and often contain additional structures such as bars and lenses \citep[][Su et al. 2020, submitted]{Janz2012,Janz2014}. Nuclei are a common feature, especially in more massive dwarf ETGs \citep{Lotz2004,Cote2006,Paudel2010,Paudel2011,Neumayer2020}. In some cluster dwarf ETGs, shells have been detected, supporting the idea that dwarf-dwarf mergers can occur inside the cluster environment \citep{Paudel2017}. Although most cluster ETGs firmly occupy the red sequence of the color-magnitude diagram \citep{Janz2009} some, which morphologically appear as dwarf ETGs, are found to have blue central regions (so called `blue-cored dwarfs', \citealp{Lisker2006b,Pak2014,Urich2017}). The variety and complexity of cluster dwarf ETGs likely suggests that there is no single origin for them, with multiple possible evolutionary tracks at play (for a review, see \citealp{Lisker2009b}).

The focus of this study is the presence of spiral arms and bar features found in some cluster dwarf ETGs. Such features were first noted in a dwarf ETG belonging to the Virgo cluster \citep[IC3328;][]{Jerjen2000}. Since then, multiple other examples have been found in Virgo \citep{Jerjen2001,Barazza2002,Ferrarese2006} and in other galaxy clusters, including Fornax \citep{DeRijcke2003,Venhola2019}, Coma \citep{Graham2003}, and Perseus \citep{Penny2014}. In general, the methods used for identifying the spiral features in these studies involved the use of unsharp masks or fitting a galaxy with a smooth model and subtracting it from the original image. \citet{Lisker2006} attempted the first systematic search for spiral features in dwarf ETGs in the Virgo cluster using Sloan Digital Sky Survey (SDSS) imaging (DR4; \citealp{Adelman-McCarthy2006}). Using an iterative technique (described in more detail in Section \ref{sec:residmeas}), \citet{Lisker2006} attempted to quantify the fraction of each dwarf ETGs light in spiral and/or bar features. Typical values ranged from $\sim$5-15\%, depending on the galaxy considered and the exact approach used to quantify it. In other words, the spiral features in dwarf ETGs were dominated by a much more luminous diffuse component. The fraction of their dwarf ETG sample presenting such features was also found to be dependent on the dwarf luminosity -- roughly half at the bright end (M$_{\rm{B}}=-17$) and falling to just a few percent at the faint end (M$_{\rm{B}}>-15$), although the exact faint end fraction is less certain due to the limited depth of the optical data.

Since the discovery of these disk-like features, a commonly discussed origin scenario is that the faint spiral features are the nearly erased remnants of low mass spiral galaxies that were morphologically transformed by environmental effects after infalling into their galaxy clusters. This interpretation was supported by the discovery that the dwarf ETGs in the cluster outskirts tend to be rotating more strongly than those in the inner cluster \citep{Lisker2009,Toloba2009}. A combination of hydrodynamical stripping \citep{GunnGott1972} and tidal interactions with the cluster potential and other cluster members \citep{Moore1998} could remove their gas content and quench their star formation, while simultaneously thickening their stellar disks to better match the observed shapes of dwarf ETGs \citep{Lisker2007,SanchezJanssen2016}. However, the discovery of isolated dwarf ETGs that are clearly rotating puts in question the need for transformation mechanisms operating in dense environments to explain the origins of rotating dwarfs \citep[][see also \citealp{Scott2020}]{Janz2017,Graham2017}. One alternative scenario could be that the rotational component was formed much earlier, prior to their infall into the cluster, perhaps during an initial primordial collapse or later grown within an initially dispersion supported system \citep{Graham2015,delaRosa2016,Tadaki2017,Wheeler2017}.

Regardless of their origin, numerous examples of rotationally supported disks in cluster dwarf ETGs have been discovered \citep{Simien2002,Geha2003,DeRijcke2003,vanZee2004,Rys2014}. In \citet{Toloba2009}, it is estimated that roughly half of their sample of dwarf ETGs are rotationally supported. Given that many of these objects orbit deep within the cluster tidal potential, it is not difficult to conceive that tidal interactions with the cluster and/or high speed encounters with cluster members might influence their disks. Several numerical simulations have shown that such tidal interactions can indeed induce bars and spiral features in cluster disk galaxies \citep{Aguerri2009,Gajda2017,Kwak2017}.

Therefore, instead of demanding that the cluster environment is necessary for forming rotationally supported disks, we investigate the cluster harassment scenario as a likely means for tidally triggering spiral and bar features in some cluster dwarf ETGs. The effects of harassment on cluster galaxies have been studied in many numerical studies -- the dependency on multiple parameters of the satellite were considered, including: surface brightness \citep{Moore1999}, halo mass and concentration \citep{Smith2010a,Smith2013a}, disk size \citep{Gnedin2003a, Smith2013a, Smith2016}, disk inclination \citep{Bialas2015}, orbit \citep{Mastropietro2005, Smith2010a, Smith2015}, and time spent in the cluster \citep{Rhee2017,Han2018}. However, there have been few dedicated studies of the dependency on stellar disk dynamics. Despite this, stellar disk dynamics likely plays a key role in dictating the response of the disk to harassment. 

For example, in \citet{Mastropietro2005}, dwarf galaxies were subjected to harassment in a cosmological context, resulting in the formation of spiral features and bars, followed by bar instabilities. Those that suffer significant stellar stripping often end up fully converted into spheroidal galaxies. However, the stellar component of their dwarf galaxies were all initialized as razor-thin disks. Such thin disks are not found in observed samples of dwarf ETGs \citep{Lisker2007}, and this choice of initial conditions may have caused their disks to be too sensitive to harassment. Thus, the spiral features generated in their model disks were much stronger than the few percent of the total luminosity found in observed cluster dwarf ETGs. Also, their final shapes are too flat and prolate (i.e., too barred) compared to observations (e.g., see Figure 12 of \citealp{SanchezJanssen2016}). Thus, it is crucial to better understand the role of a disk's stellar dynamics in dictating the response to harassment. This is a key motivation for our study. 

Using our controlled simulations of cluster tides derived from a cosmological simulation, we can subject model dwarf ETGs to identical tidal histories while systematically varying the degree of rotational support within their stellar disk. In this way, we can cleanly determine the role of their stellar disk dynamics. As we demonstrate in this study, we find that the degree of rotational support does not strongly affect the amount of stellar mass loss or the amount of disk thickening. However, we find there are strong consequences for the formation of spiral- and bar-like features that require a much higher degree of rotational support than is typically observed. We interpret this result as implying the presence of a highly rotationally supported disk component, embedded in a more dispersion supported component, which dominates the light and mass of the stellar disk. Although the origin of this highly rotationally supported component is not clear (see Section \ref{sec:discussion} for further discussion of possible origins), we find it can be effectively tidally triggered to form spiral and bar features that appear remarkably similar to those we observe.

This work is organized as follows. In Section 2 we describe our sample, the method for measuring their residuals, and present the resulting residual fractions. In Section 3 we explain our numerical simulations set-up. In Section 4 we present our simulation results, and we discuss our results in Section 5. Finally, in Section 6 we summarise and draw conclusions.

\section{Observational Data}
\label{obsdatasect}

\subsection{The Sample}
In this study, we focus on the nine Virgo cluster passive dwarfs in our companion paper (Brought to Light I: Michea et al. 2021). These are VCC 0216, 0308, 0490, 0523, 0856, 0940, 1010, 1695, and 1896. The dwarfs were originally selected to be certain cluster members, and were already known to show evidence for spiral arms in SDSS images \citep{Lisker2006b}. The dwarfs were then observed with the ESO 2.2m Wide Field Imager (WFI; program 077.B-0785). A white filter was used to maximize the signal, with a net exposure time of 2.5 hours in the optimal case (for some targets, this was not achieved due to bad weather). The final images reach a typical r-band surface brightness of $\sim$27.8 mag arcsec$^{-2}$, which would correspond to 29 mag arcsec$^{-2}$ in B-band assuming typical early-type dwarf colors, at a signal-to-noise (S/N) ratio of 1 per binned pixel (0.71\arcsec). The actual depth reached for individual different targets is given in the third column of Table \ref{obssample_tab}. For a more detailed description of the data and how they were processed, please see our companion paper (Brought to Light I: Michea et al. 2021).

\begin{table}[]
    \centering
    \caption{Columns from left to right; galaxy name, galaxy luminosity, image depth, and summary of main residual features and the residual light fraction $f_{\rm{residual}}$ .}
    \begin{center}
    \begin{tabular}{|c|c|c|c|l|c|}
    \hline
       VCC  & R$_{\rm{abs}}$ & Depth$^\alpha$ & Features$^\beta$ & $f_{\rm{residual}}^\delta$\\
       \hline
        0216 & -16.67 & 27.94  & spiral & 0.029$^{+0.004}_{-0.002}$  \\
        0308 & -17.79 & 27.75  & spiral & 0.024$^{+0.002}_{-0.002}$ \\
        0490 & -18.03 & 27.91  & spiral, bar & 0.036$^{+0.004}_{-0.003}$ \\
        0523 & -18.34 & 27.21  & spiral, bar, LP?$^\gamma$ & 0.042$^{+0.003}_{-0.004}$ \\
        0856 & -17.62 & 27.89  & spiral & 0.022$^{+0.004}_{-0.002}$ \\
        0940 & -17.12 & 27.59  & spiral, bar & 0.035$^{+0.003}_{-0.003}$ \\
        1010 & -18.22 & 27.66  & spiral, bar, LP?$^\gamma$ & 0.030$^{+0.004}_{-0.005}$ \\
        1695 & -17.48 & 27.86 & spiral, bar & 0.054$^{+0.005}_{-0.007}$ \\
        1896 & -16.84 & 27.79 & spiral, bar & 0.064$^{+0.011}_{-0.013}$ \\
        \hline
    \end{tabular}
    \newline 
    \newline $^\alpha$ Surface brightness depth in r-band at a $S/N$ = 1 and a scale of 0.71 arcsec pixel$^{-1}$.
    \newline $^\beta$ Broad description of observed features based on Fourier mode analysis (see companion paper).
    \newline $^\delta$ Residual light fraction measured within two effective radii.
    \newline $^\gamma$ Possible enhancements at Lagrangian points.
    \end{center}
    \label{obssample_tab}
\end{table}

\begin{figure*}
\makebox[\textwidth]{\includegraphics[width=170mm]{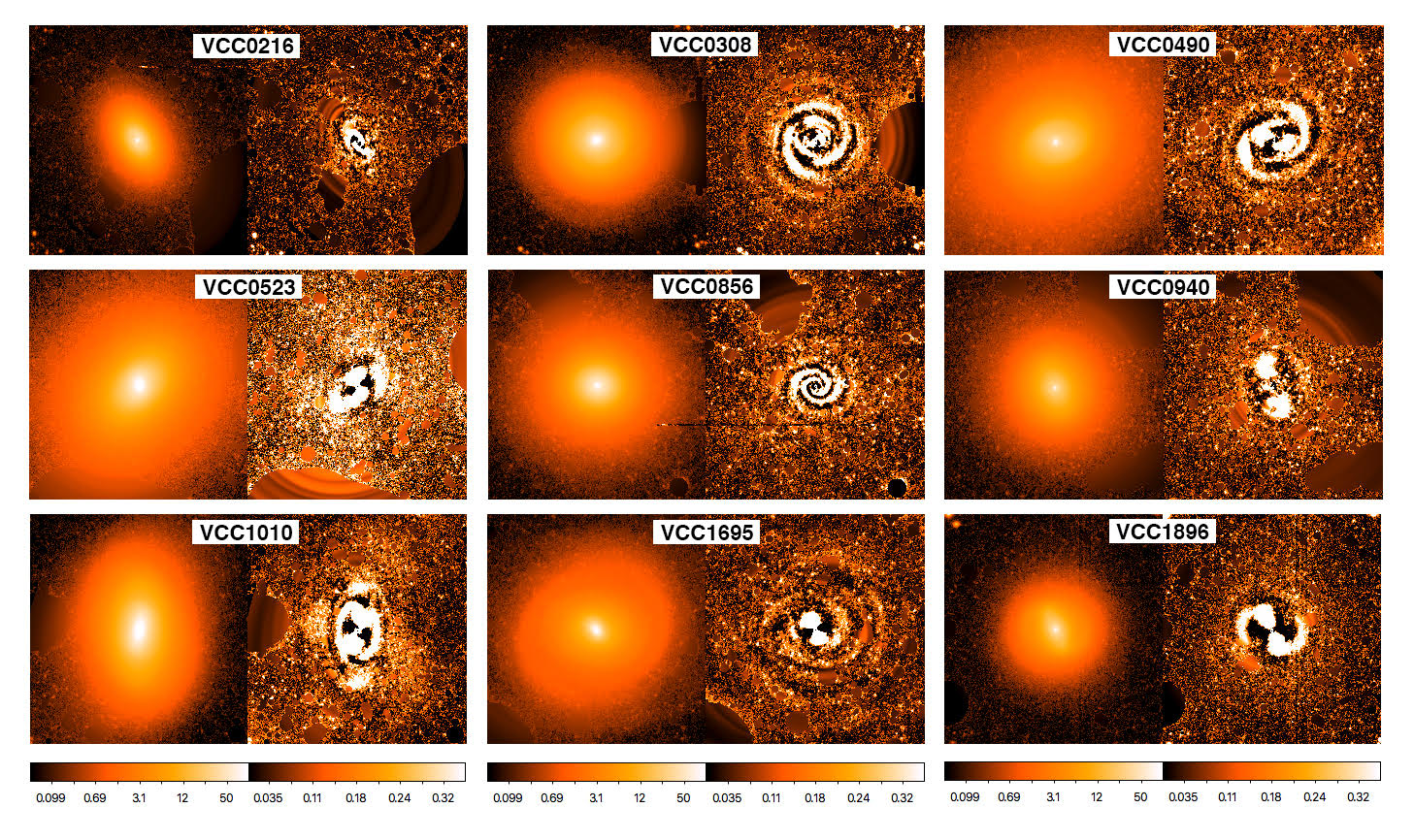}}
\caption{Left panels: Deep imaging of our sample of passive dwarf galaxies from the Virgo cluster. Right panels:
Residual images after subtracting off the diffuse model component.
The panels are 65 by 55 arcseconds, the color-bar units are background subtracted flux. These images can also be found in our companion paper (Brought to Light I: Michea et al. 2021).}
\label{residuals_sample}
\end{figure*}

\begin{figure*}
\includegraphics[width=\textwidth]{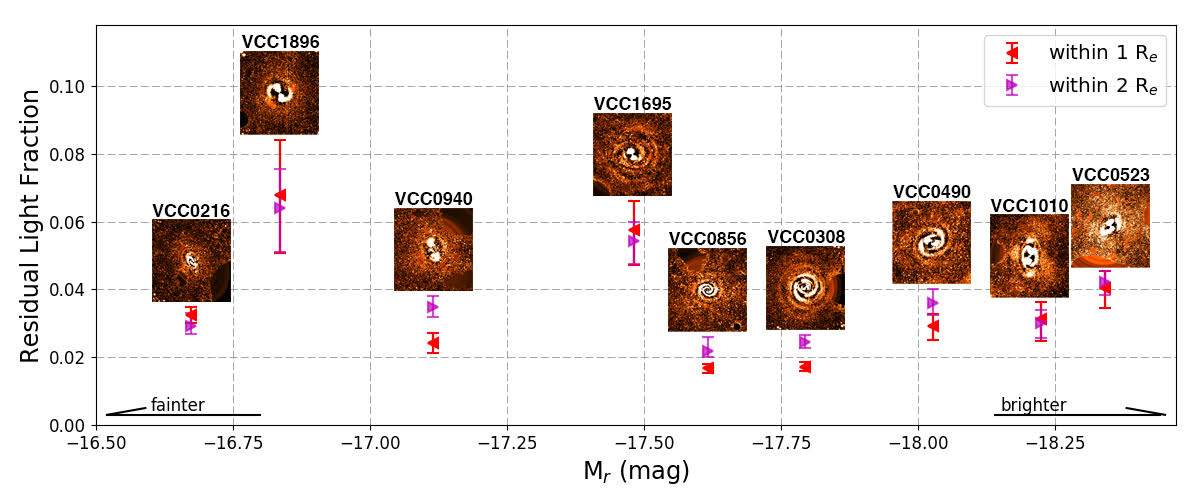}
\caption{Results of the residual imaging procedure on our
observed sample of passive dwarf galaxies. The Y-axis is the `residual light fraction', which is the fraction of the total galaxy light that appears in the residual maps, measured within one or two effective radii (see legend), 
compared to the total r-band absolute magnitude. This figure is adapted from Figure 5 of our companion paper (Brought to Light I: Michea et al. 2021).}
\label{residualresultsfig}
\end{figure*}

\subsubsection{Method for Revealing Residual Features}
\label{sec:residmeas}
A full description of our methodology for producing residual maps and associated testing is presented in our companion paper (Brought to Light I: Michea et al. 2021). Here we provide a short, concise description of the method for completeness.

Our method is originally based on an iterative technique put forward in \cite{Lisker2006} to separate out the diffuse component from the substructure component, but with some significant and substantial improvements. In short, the method attempts to fit the diffuse component of a galaxy using the {\sc{iraf}} task {\sc{ellipse}}, in conjunction with a smoothing procedure. The diffuse component is then subtracted from the original image to create a residual image. However, this typically results in over-subtraction, creating negative pixels in interarm regions of the residual map. So, now a copy of the residual map is made but with all negative pixels set to zero. This is then subtracted from the original image, which reduces the substructure present in it. The process then begins again using this new image with the reduced substructure, and is repeated iteratively, gradually removing the substructure from the image. We exit the iterative loop once the negative pixels in the interarm region have reached a median value of zero, which is one of several key improvements over the \cite{Lisker2006} approach (see companion paper, Brought to Light I: Michea at al. 2021 for more details). The end result provides an ellipse model with radially varying ellipticity and position angle for the diffuse component, and a residual map of the substructure.

\subsubsection{Results of New Method on Our Sample}
We apply this technique to our full sample of nine galaxies. The original image (left) and residual map (right) is shown for each galaxy in Figure \ref{residuals_sample}. While some hint of structure is visible in the original images, the residual method highlights often spectacular spiral-like features. For instance, VCC308 and VCC856 present a clear set of tightly wound spiral features. Meanwhile several of the galaxies present prominent bars (e.g., VCC0490, VCC0523, VCC0940, VCC1010, VCC1695, VCC1896), often with bold spiral features protruding from the ends of the bar (VCC0523, VCC1010, VCC1896). In VCC0523 and VCC1010 there are luminous spots near to but not directly connected with the bar. These could perhaps represent the Lagrangian points due to the non-axisymmetric shape of the potential arising from the bar. In our companion paper (Brought to Light I: Michea et al. 2021) we quantitatively classify these spiral features and their bars using a Fourier analysis. Here we include only a brief summary of the main morphological features in Table \ref{obssample_tab}.

In order to quantify the amount of light present in disk features such as these, we measure the total flux within the residual image and in the original image, and divide them to calculate the {\it{residual light fraction}}. We measure the residual light fraction within one and two effective radii. The results for each dwarf galaxy are shown in Figure \ref{residualresultsfig} as a function of their r-band absolute magnitude. The residual light fractions vary between 2.2-6.4\% within two effective radii. There is no clear trend between the residual light fraction and the galaxy luminosity. We also provide the residual fractions in the right-most column of Table \ref{obssample_tab}.

\section{Numerical Simulations}

\subsection{The Code}
In this study we make use of `{\sc{gf}}' (\citealp{Williams2001}; \citealp{Williams1998}), which is a TREE-SPH algorithm that operates primarily using the techniques described in \cite{Hernquist1989}. The early-type dwarf galaxies we consider are gas-free and therefore not star-forming. We therefore do not include an SPH component or star formation recipe in our models, and so we operate the code purely as a gravitational code without considering hydrodynamics. `{\sc{gf}}' has been parallelized to operate simultaneously on multiple processors to decrease simulation run-times. The Treecode allows for rapid calculation of gravitational accelerations. In all simulations, the gravitational softening length, $\epsilon$, is fixed for all particles at a value of 25 pc. In practice, we find this is sufficient to suppress numerical scattering between star particles, or of star particles off dark matter particles. We must have a high spatial resolution to ensure we can resolve well the disks of dwarf galaxies, and structures that form within them. Gravitational accelerations are evaluated to quadrupole order, using an opening angle $\theta_c=0.7$. A second order individual particle timestep scheme was utilized to improve efficiency following the methodology of \cite{Hernquist1989}. Each particle was assigned a time-step that is a power of two division of the simulation block timestep, with a minimum timestep of $\sim 0.5$ yrs. Assignment of time-steps for collisionless particles is controlled by the criteria of \cite{Katz1991}. For details of code testing, please refer to \cite{Williams1998}.

\subsection{The Harassment Model}

\begin{figure*}
\includegraphics[width=180mm]{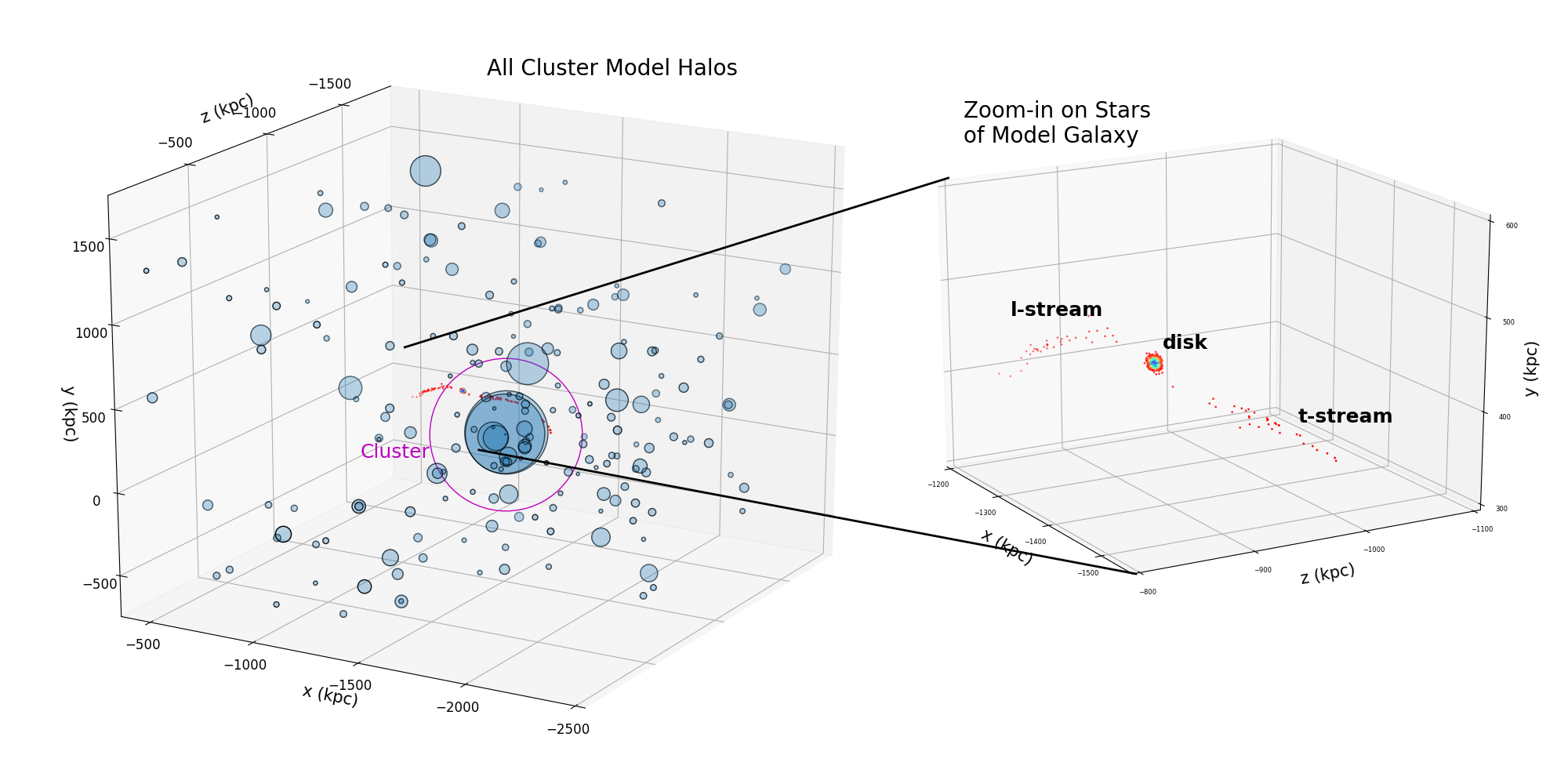}
\caption{Left panel: 3D visualization of the simulated cluster, with an orbiting dwarf galaxy
model. Blue filled circles show the dark matter halos of the tidal field of the cluster. The main
cluster halo is shown with a purple circle. The circle size is equal to the outer radius of the
halos. Rainbow-colored points show the surface density of the stellar disk of a harassed model galaxy. Right panel: a zoom-in box on the model galaxy’s stars, highlighting the leading (l-stream) and trailing (t-stream) tidal streams. The streams are shown in red as they are low surface brightness compared to the stellar disk.}
\label{3dhalodistrib}
\end{figure*}

Following the approach of \cite{Smith2015}, we model the dynamical and time-evolving potential of a galaxy cluster using analytical potentials. An analytical gravitational potential is used for the main cluster, and each individual `harasser galaxy' (cluster members that can interact with a model galaxy) has its own unique analytical potential. This approach has advantages -- the spatial resolution of gravity from harasser galaxies is effectively infinite. Also placing high resolution model galaxies in cosmological simulations (e.g., \citealp{Moore1999}; \citealp{Mastropietro2005}) is numerically challenging. In comparison, computing accelerations from analytical potentials is very fast. However, there are also disadvantages -- harasser galaxies move on fixed tracks and cannot respond to the gravity of the live model galaxy. Fortunately this is of negligible consequence for modeling the high speed tidal encounters that typically occur between cluster members and, to be safe, we deliberately choose orbits that result in high-speed encounters only.

The properties of the main cluster halo and harasser galaxy halos are dictated by a cosmological simulation of a cluster. We shall refer to the main cluster halo as the `cluster halo' hereafter, in order to distinguish it from other halos. For each snapshot of the cosmological simulation we measure the properties (e.g., virial mass and radius) of all halos at that instant. Then for each halo we construct an analytical potential that mimics each individual halo's tidal field. By applying the tidal field from all halos simultaneously, we construct the total tidal field of the cluster. In this way, we allow for the growth in mass and structural evolution of the cluster with time as it accretes new galaxies, and the properties of individual harasser galaxies also evolve in time.

 The original cosmological N-body simulation on which our harassment model is based on was performed with the adaptive mesh refinement code {\sc{MLAPM}} (\citealp{Knebe2001}), and is fully described in \cite{Warnick2006} and \cite{Warnick2008}. In the cosmological simulation, each dark matter particle has a mass $\sim1.6\times10^8 h^{-1} $M$_\odot$, and the highest spatial resolution is $\sim 2$ kpc. The cluster used is C3 from Table 1 of \cite{Warnick2006}. At $z=0$, the cluster has a virial mass of 1.1$\times$10$^{14} h^{-1} $M$_\odot$, and a virial radius of 973 $h^{-1}$ kpc. The cluster is followed for $\sim 7$ Gyr (since $z=0.8$), during which time it approximately doubles in mass. A snapshot is produced once every $\sim 0.16$ Gyr. This high time resolution enables us to follow the orbits of individual halos with high accuracy. In any snapshot, halo properties are measured with the halo finder MHF (MLAPM's halo finder; \citealp{Gill2004a}), down to 20 particles per halo. Therefore the minimum resolved halo mass is $\sim3 \times 10^9 h^{-1}$ M$_\odot$. There are a total of 402 halos including the cluster and harasser halos.

In our harassment model, each halo's potential well is described by a Navarro, Frenk and White (NFW) analytical potential (\citealp{Navarro1996})

\begin{equation}
\label{potNFW}
\Phi = -g_cGM_{200}\frac{\ln(1+(r/r_s))}{r}\rm{,}
\end{equation}
\noindent where $g_c=1/[{\ln(1+c)-c/(1+c)}]$, $c$ is the concentration parameter that controls the shape of the profile, $r_{\rm{s}}$ is a characteristic radial scalelength, and $M_{200}$ is the virial mass.

\begin{figure*}
\makebox[\textwidth]{\includegraphics[width=170mm]{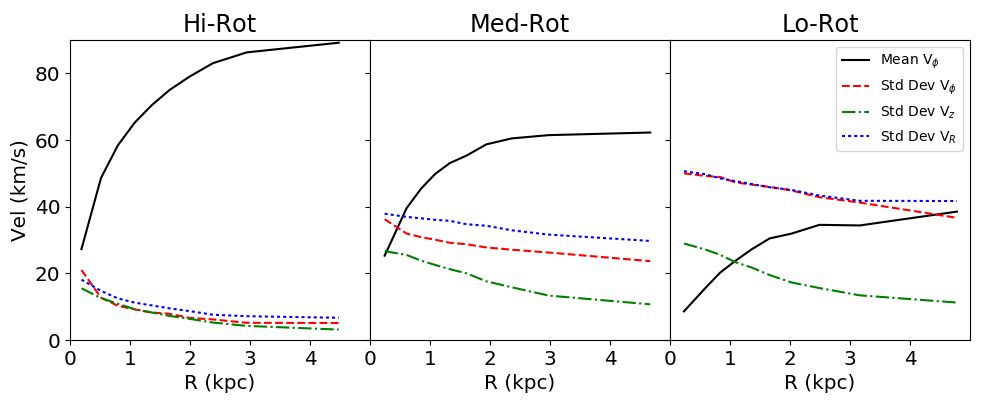}}
\caption{Velocity profiles of model galaxy initial conditions. Shown from left to right in order of decreasing rotational support, the `Hi-Rot' model (fully rotationally dominated at all radii), the `Med-Rot' model, and the `Lo-Rot' model (dispersion dominated at all radii). The legend indicates the dynamical measurement.}
\label{ICs_diskdyn}
\end{figure*}

The properties and position of each halo in a snapshot are taken directly from the cosmological simulation using the \cite{Gill2004a} halo tracker. Between snapshots, we linearly extrapolate halo positions and properties. As the original snapshots are only $\sim 0.16$ Gyr apart, this leads to smooth halo motion as a function of time. This procedure allows us to calculate the potential and accelerations at any position, and at any moment over the $\sim 7$ Gyr duration of the simulation.

In Figure \ref{3dhalodistrib}, we illustrate the appearance of the cluster model's halo distribution at $z=0$. Each transparently filled blue circle represents a dark matter halo with size equal to the virial radius. The main cluster halo is outlined with a purple empty circle. We overlay the long tidal tails produced by the stellar component of the thin model on Orbit S1. The right-hand panel shows a zoom in on just the stars of the model disk, revealing long tidal streams that lead and trail the galaxy as it orbits within the cluster potential.

\subsection{Dwarf Galaxy Models}
\label{sec:models}

Our model galaxies consist of 2 components; a NFW dark matter halo, and a thick exponential distribution of stars. \cite{Smith2013a} have already completed a detailed parameter study of how the effects of harassment depend on multiple galaxy properties including halo mass, halo concentration, disk size and mass, and globular cluster population. Therefore, in this study we do not vary these properties. Instead, here we concentrate on how the effects of harassment depend on one main parameter -- the amount of rotational support present in the stellar dynamics of the disk.

\vspace{2mm}
\noindent {\it{Dark matter halo: }}The dark matter halo of our galaxy models has an NFW density profile
\begin{equation}
\rho(r) = \frac{\delta_{\mathrm{c}} \rho_{\mathrm{crit}}}{(\frac{r}{r_{\rm{s}}})(1+\frac{r}{r_{\rm{s}}})^2}\rm{,}
\label{NFWdensprof}
\end{equation}
\noindent where $\delta_{\mathrm{c}}$ is the characteristic density, and $\rho_{\mathrm{crit}}$ is the critical density of the Universe. Given that the NFW model has a divergent total mass, we truncate the profile at $r_{\rm{200}} = r_{\rm{s}} c$ where $c$ is the halo concentration. Our model galaxy has a dark matter halo mass of $10^{11}$M$_\odot$, consisting of 100,000 dark matter particles, with a concentration $c = 14$. The virial radius is $r_{\rm{200}} = 95$~kpc with respect to the field. The peak circular velocity of the halo is 88~km~s$^{-1}$ at a radius of 15~kpc.

\vspace{2mm}
\noindent {\it{Stellar disk: }}The stellar distribution of our model galaxies has a radially exponential form (\citealp{Freeman1970})
\begin{equation}
\label{expdisk}
\Sigma(R) = \Sigma_0 \exp (R/R_{\rm{d}})\rm{,}
\end{equation}

\noindent
where $\Sigma$ is the surface density, $\Sigma_{\rm{0}}$ is central surface density, $R$ is radius within the stellar distribution, and $R_{\rm{d}}$ is the exponential scalelength.

The size and mass of the stellar distribution is chosen to approximately match the observed properties of luminous Virgo early-type dwarfs. These are characterized by close-to-exponential luminosity profiles (\citealp{Janz2008}). Our model has a total stellar mass of $3.0 \times 10^{9}$M$_\odot$ ($3\%$ of the halo mass; \citealp{Peng2008}), and is formed from 100,000 star particles. We aim to have a stellar distribution with an effective radius $R_{\rm{eff}}=1.5~$kpc in all our models. This is consistent with the sizes of early-type galaxies with this stellar mass from \cite{Janz2012}, and typical for our sample of dwarfs (see Table 2 of our companion paper; Brought to Light I: Michea et al. 2021). We try to keep this value fixed between all our models because the ease with which tidal stripping of stellar disks occurs is known to be a strong function of the ratio of the stellar disk size to the halo size \citep{Smith2016}, and we are more interested in the effects of varying stellar disk dynamics in this study, as described below.

\vspace{2mm}
\noindent {\it{Varying stellar disk dynamics: }}We consider three main model galaxies, between which we vary their stellar disk dynamics. Their velocity profiles are shown in Figure \ref{ICs_diskdyn}. 

The `Hi-Rot' model is a highly rotationally supported disk. Rotational motion easily dominates at all radii. The opposite extreme is the `Lo-Rot model' which is dominated by dispersion at all radii. The `Med-Rot' model is intermediate between the two in terms of its rotational support. 

The `Hi-Rot' model is thin, and half of its stars can be found within a distance of 43~pc from the plane of the disk (denoted $z_{1/2}=43$~pc), corresponding to a vertical-to-radial axis ratio of about 0.1. The latter two models have thicker disks with $z_{1/2}=180$~pc, corresponding to a vertical-to-radial axis ratio of 0.4-0.5, which is a good match to the mean vertical-to-radial axis ratio of Virgo cluster early-type dwarfs \citep{Ruben2016}.

\vspace{2mm}
\noindent {\it{Comparison with observed stellar disk dynamics: }}The degree of rotational support in observed cluster dwarfs is typically found to be in the range of V$_{\rm{peak}}$/$\sigma_0=0.05$-1.0 and, in a few rare cases as high as $\sim1.5$ or more \citep{Zee2004b, Toloba2009, Rys2014}. Therefore, most observed early-type dwarfs have less rotational support than the 'Med-Rot' model, which has V$_{\rm{peak}}$/$\sigma_0=1.1$, and some have less rotational support than the `Lo-Rot' model V$_{\rm{peak}}$/$\sigma_0=0.5$. Few (if any) observed early-type dwarfs reach the high rotational support of the `Hi-Rot' model (V$_{\rm{peak}}$$/$$\sigma_0=2.8$). Thus, the `Hi-Rot' model should be considered quite extreme in terms of its rotational support (and also its shape, as described in the previous paragraph) compared to observed early-type dwarfs. However, the `Hi-Rot' is crucial to our study as we wish to study the response of a highly rotationally supported disk to harassment. And, as we will demonstrate later, the tidally induced structures that such extreme disks produce bear a striking resemblance to those we detect in the residual maps of real early-type cluster dwarf galaxies.

\vspace{2mm}
\noindent {\it{Set-up code: }}Positions and velocities of dark matter and star particles are assigned using the publically available initial conditions set-up code {\sc{DICE}} \citep{Perret2014}. We find this produces initial conditions that are already very stable. However, for additional stability, we initially evolve our models in isolation for 0.5 Gyr to ensure that they are fully relaxed. Then we introduce them into our harassment model.

\subsection{Choice of Orbits}
\label{sec:orbits}
\begin{figure}
\includegraphics[width=\columnwidth]{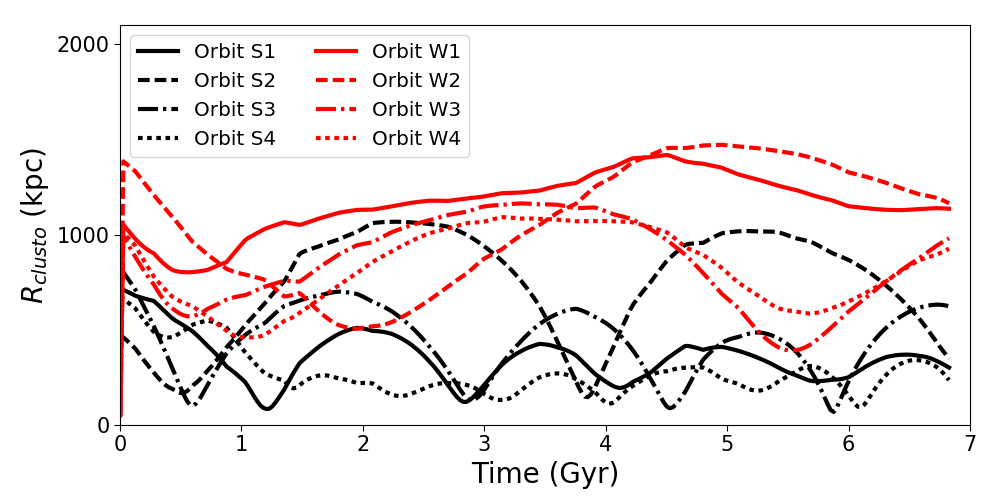}
\caption{Evolution of the clustocentric radii of galaxies along the orbits we consider. Strong harassment orbits are shown in black (orbit S1-S4) and weak harassment orbits are shown in red (orbit W1-W4). The weak harassment orbits can be considered quite typical. But, the strong harassment orbits have large numbers of close pericenter passages, meaning they should be considered as examples of unusually strong harassment (see text for more details).}
\label{Orbitcompfig}
\end{figure}

In \cite{Smith2015}, we considered how the effects of harassment depend on the orbit by modeling over 168 individual orbits, spanning a wide range of ellipticities and pericenter distances. Having previously studied orbital dependence in detail, here we select only 8 orbits from the full 2015 set. We select four `strong harassment' orbits and four `weak harassment' orbits. These are illustrated in Figure \ref{Orbitcompfig}. 

The `strong harassment' orbits (black curves) have much smaller pericenter distances and tend to have multiple pericenter passages, compared to the `weak harassment' orbits. We compare their pericenter distances and number of pericenter passages with that of the distribution found in the \cite{Smith2015} study. The `strong harassment' orbits have systematically smaller pericenters (at the one-sigma level), and systematically larger numbers of pericenter passages (at a greater than two-sigma level) than the mean values of the 2015 simulation orbital set. Thus the `strong harassment' orbits should be considered as quite extreme cases of strong harassment. In comparison, the `weak harassment' orbits should be considered much more typical, as almost all the points lie within one sigma of the 2015 mean value for both pericenter distance and number of pericenter passages.

One significant advantage of our harassment model over using a full cosmological simulation is that we are able to place a chosen galaxy model on exactly the same orbital trajectory through the cluster. In this way, each model galaxy is subjected to an identical tidal history, and any differences between the models are purely as a result of differing internal properties of the models. Thus, we can take full advantage of the controlled nature of our simulations to undertake a systematic study of how stellar disk dynamics alters a galaxy's response to harassment. Such a precise analysis of an individual parameter (e.g., rotational support) is very challenging in a cosmological simulation, as each galaxy will have differing orbits and thus differing tidal histories, as well as varying internal properties (such as mass, concentration, stellar dynamics, merger history, etc). We subject each model galaxy (`Hi-Rot', `Med-Rot', and `Lo-Rot') to the four strong harassment orbits (Orbit S1-S4), and we additionally subject the `Hi-Rot' model to the four weak harassment orbits (Orbit W1-W4). For reference, these are listed in Table \ref{tab:finalboundfractions}.

\section{Results}

\subsection{Tidal Mass Loss}
\label{sec:massloss}
\begin{table}[]
    \centering
    \begin{tabular}{|c|c|c|c|c|c|c|}
    \hline
    Model & Orbit & $R_{\rm{eff},\rm{i}}$ & $z_{\rm{i}}$ & z$_{\rm{f}}$ & $f_{\rm{dm}}$ & $f_{\rm{str}}$   \\
     & & (kpc) & (pc) & (pc) & & \\
    \hline
    \multirow{5}{*}{Hi-Rot} & isol & \multirow{5}{*}{1.51} & \multirow{5}{*}{43} & 106 & 1.000 & 1.000 \\
         & W1 &      &    & 123 & 0.628 & 1.000 \\
         & W2 &      &    & 122 & 0.533 & 1.000 \\
         & W3 &      &    & 124 & 0.409 & 1.000 \\
         & W4 &      &    & 130 & 0.454 & 1.000 \\   
    \hline
    \multirow{5}{*}{Hi-Rot} & isol & \multirow{5}{*}{1.51} & \multirow{5}{*}{43} & 106 & 1.000 & 1.000 \\
         & S1 &      &    & 192 & 0.091 & 0.994 \\
         & S2 &      &    & 140 & 0.221 & 1.000 \\
         & S3 &      &    & 186 & 0.071 & 0.997 \\
         & S4 &      &    & 232 & 0.046 & 0.933 \\
    \hline
    \multirow{5}{*}{\shortstack[c]{Med-Rot}} & isol & \multirow{5}{*}{1.46} & \multirow{5}{*}{196} & 241 & 1.000 & 1.000 \\
         & S1 &      &    & 314 & 0.091 & 0.985 \\
         & S2 &      &    & 284 & 0.218 & 1.000 \\
         & S3 &      &    & 353 & 0.070 & 0.980 \\
         & S4 &      &    & 414 & 0.046 & 0.918 \\
    \hline    
    \multirow{5}{*}{\shortstack[c]{Lo-Rot}} & isol & \multirow{5}{*}{1.50} & \multirow{5}{*}{187} & 248 & 1.000 & 1.000 \\
         & S1 &      &    & 374 & 0.090 & 0.960 \\
         & S2 &      &    & 330 & 0.217 & 0.996 \\
         & S3 &      &    & 424 & 0.069 & 0.937 \\
         & S4 &      &    & 615 & 0.045 & 0.864 \\
    \hline
    \end{tabular}
    \caption{Galaxy models and orbits, initial $R_{\rm{eff}}$, the initial and final vertical thickness (the vertical height containing half the star particles), the final bound dark matter fraction and final bound star fraction.}
    \label{tab:finalboundfractions}
\end{table}

In Figure \ref{masslossfig}, we plot the remaining bound fraction of dark matter (f$_{\rm{dm}}$; x-axis) and the remaining bound fraction of stars (f$_{\rm{str}}$; y-axis) at the end of the simulation. These results are also listed in Table \ref{tab:finalboundfractions}. From left to right, the five data points correspond to the bound fractions for Orbit S4, Orbit S3, Orbit S1, Orbit S2, and the final points on the right with f$_{\rm{dm}}=1$ are the `isolated' control models, which are evolved in isolation for the same duration as the harassment models. Thus, the strongest mass loss occurs for the left-most data points (Orbit S4), in which only 4.5\% of the halo remains bound (f$_{\rm{dm}}=0.045$) at $z=0$. We exclude the weak harassment orbits from this plot for clarity, but they would all be found on the near horizontal line in the upper-right corner of the panel.

Figure \ref{masslossfig} allows us to see how much stellar stripping  occurs as the dark matter halo is stripped. Only when very large amounts of dark matter are stripped (f$_{\rm{dm}}$ reaches $\sim 0.1$) do we see any significant decrease in the stellar mass fraction. Although we only consider `strong harassment' cases in this plot, which we note are quite extreme in Section \ref{sec:orbits}, none of our models lose significant quantities of stars, with the maximum mass loss resulting in f$_{\rm{str}}$ being reduced from 1 to 0.86.

The sample is split by the rotational support of the model galaxy (see legend). At first glance, there appears to be slightly more stellar mass loss for more dispersion supported disks. However, when we measure $R80$, which is the radius from the disk center containing 80\% of the stars in the pre-harassment disks, we find that the outer disks of the more dispersion supported models are slightly more extended (see $R80$ values in legend). We deliberately matched their effective radii when setting up the disks (see Table \ref{tab:finalboundfractions}). But we relax the initial conditions in isolation before introducing the models into the harassment model. And during the relaxation process, it appears the outermost disk stars end up slightly more extended in the more dispersion supported models (although the models still match in effective radius). Therefore, some of the increased stellar mass loss must be due to their more extended stellar disks. In any case, the difference in stellar mass loss between the models is quite mild for a fixed amount of halo mass loss (just a few percent), so any dependency on the internal dynamics must be minor.

In summary, our `strong harassment' orbits drive significant mass loss from the dark matter halos of our models. But even for these quite extreme orbits, our models do not lose significant quantities of stars due to the fact that the stellar disk is much smaller than their dark matter halos. The small differences between the amount of stars stripped can be mostly explained by slight differences in the size of their outer stellar disks. Therefore, we see no evidence that the amount of stellar mass lost is highly sensitive to their stellar disk dynamics.

\subsection{Disk Thickening}
\label{sec:thickening}
We quantify the thickness of our galaxy models using the parameter $z_{1/2}$, defined as the distance out of the plane of the disk, in both directions, that contains half the stars of the disk. As harassment can strip stars into long stellar streams that could artificially enhance our measurement of the disk thickness, we only measure $z_{1/2}$ within a 6 kpc radius of the model galaxy center.

\begin{figure}
\includegraphics[width=\columnwidth]{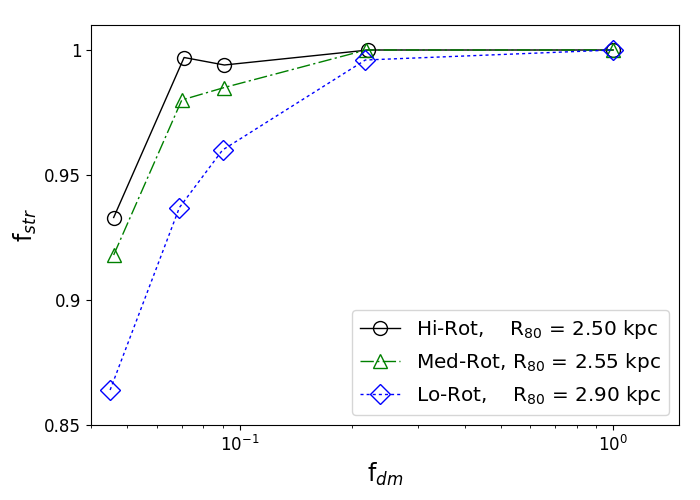}
\caption{Remaining bound fraction of stars vs dark matter after harassment. As models lose dark matter, they move from right to left, and as they lose stars they move from top to bottom. Although more dispersion supported systems appear to lose more mass, this is likely due to the fact their outer disks are slightly more extended, as measured using $R80$, the radius containing 80\% of the disk stars in the pre-harassment disks (see legend). We exclude the `weak harassment' orbits from these plots for clarity.}
\label{masslossfig}
\end{figure}

It is well known that stellar disks in numerical simulations suffer artificial thickening as a result of scattering of star particles off other particles. To account for this effect, we measure the amount of disk thickening in each model relative to a control model. The control model is the same model galaxy, but evolved in isolation and for the same duration as the models are evolved in the harassment model. 

\begin{figure}
\includegraphics[width=\columnwidth]{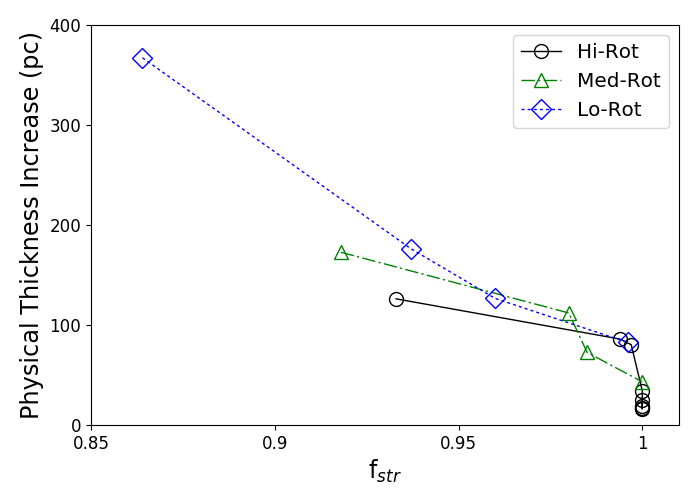}
\caption{Disk thickening due to harassment measured with respect to the isolated control counterpart as a function of the final bound stellar fraction. The disk thickness is measured as the distance from the disk plane (in both directions) containing half the disk stars, $z_{1/2}$. All models follow a similar trend for increasing physical thickness with increasing stellar stripping, independent of their rotational support. Also, there is little thickening of the disk until stellar stripping begins (after $\sim90\%$ of the halo is stripped, see Figure \ref{masslossfig}).}
\label{diskthickfig}
\end{figure}

In Figure \ref{diskthickfig}, we plot the physical increase in thickness (measured as the difference in thickness between the harassed model and the isolated control model) on the y-axis, versus the final bound stellar fraction $f_{\rm{str}}$. Most of the models follow a single trend for increasing disk thickening, but there is little thickening until stars begin to be stripped from the disk. Figure \ref{masslossfig} shows that this does not occur until the dark matter halo has already been heavily stripped and lost $\sim$80-90\% of its bound mass. So, in short there is very little stellar disk thickening until our models have lost a substantial amount of their total mass. 

\begin{figure*}
\makebox[\textwidth]{\includegraphics[width=150mm]{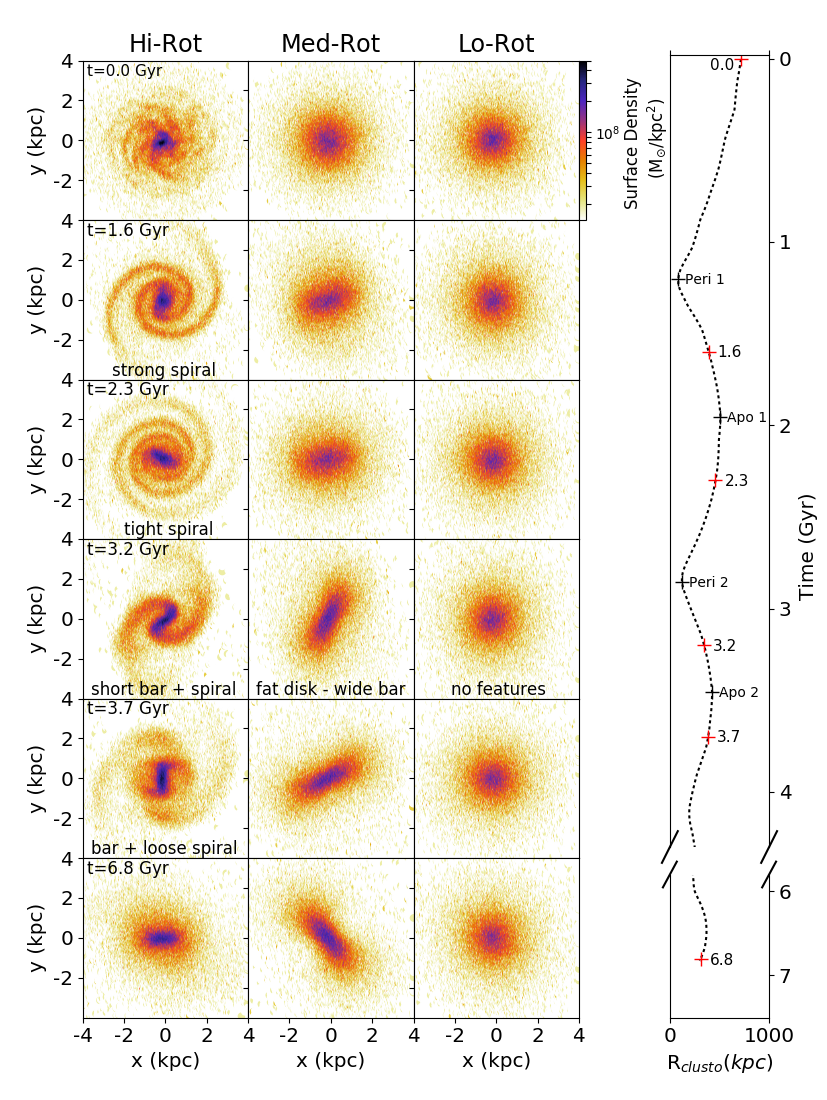}}
\caption{Tidally induced features in simulated galaxy disks and dependency on disk dynamics. Surface density plots of the face-on model disks, following the Orbit S1. Left to right
columns show the model galaxy (labeled at the top of each column), in order of decreasing rotational support. Top to bottom shows an increasing time of the snapshot, with the time given in the upper-left corner of the first column. The vertical plot shows clustocentric radial evolution with time, and red crosses marking the instants of the snapshots. Black crosses indicate the moments of pericenter and apocenter passages.}
\label{diskresponsefig}
\end{figure*}

We note that if we had plotted the fractional increase in the disk thickness, a physical increase of $\sim$100~pc would be much more significant for the `Hi-Rot' model, which is much thinner than the other models. But, by plotting the physical increase, we highlight that the true physical changes in disk thickness in response to harassment are not highly sensitive to their initial internal stellar dynamics.

Summarizing, there is negligible disk thickening from harassment until the dark matter halo has been heavily truncated, and stars begin to be stripped. At this point, in our models the physical disk thickness increases with the fraction of stars stripped. As with tidal mass loss, we do not see a strong dependency in the amount of disk thickening on stellar disk dynamics.

\subsection{Tidally Induced Features}
\label{sec:inducedfeatures}
\subsubsection{Dependency on Disk Dynamics}

So far, we have seen that the tidal mass loss of stars and stellar disk thickening do not show a strong dependency on stellar disk dynamics (see Sections \ref{sec:massloss} \& \ref{sec:thickening}). However, as we will now show, this is certainly not the case for the morphological response of the disk. We find that the strength of tidally induced features such as bars and spiral arms is highly dependent on the degree of rotational support within the disks.

In Figure \ref{diskresponsefig}, we present time-evolution (from top to bottom) of stellar-disk surface-density maps of the three 
model galaxies (see column header). The first row shows the disks at t=0~Gyr. In this figure, all the models follow the same orbit (Orbit S1), however similar responses to tidal triggering are seen for the other strong harassment orbits too (i.e., Orbit S2-S4). The red crosses in the vertical plot on the right also illustrate where they are in their orbit at the instant of the snapshots.

\vspace{2mm}
\noindent {\it{After first pericenter passage (t=1.6~Gyr):}} The second row is shown shortly after first pericenter passage. It is clear that the high rotational support model (`Hi-Rot') generates clear spiral structure in response to the tidal shock while passing the cluster pericenter. It also generates a compact bar. We note that tidally induced features such as these are only generated just after cluster pericenter passage. This suggests that tides from the main cluster potential are a primary origin for such features, although galaxy-galaxy encounters could also play a role as such encounters occur more frequently near the cluster center. In contrast to the `Hi-Rot' model, after the first pericenter passage, the `Med-Rot' and `Lo-Rot' models remain smooth and featureless, except for a slight elongation of the stellar disk of the 'Med-Rot' model.

\vspace{2mm}
\noindent {\it{After first apocenter (t=2.3~Gyr):}} 
In the `Hi-Rot' model, bars and spiral arm features that were generated shortly after the first pericenter passage are still clearly visible even after passing the first apocenter. This means these tidally induced features last for $\sim1-2$~Gyr, and actually are often still present at the time of the second pericenter. This is important because it means that, despite being generated at pericenter passage, tidally induced features can be present over a large range of clustocentric radii, which acts to flatten any radial dependencies within the cluster. However, during this time the features evolve, with the spiral arms becoming increasingly more tightly wrapped.

\vspace{2mm}
\noindent {\it{After second pericenter (t=3.2~Gyr):}} 
The second pericenter drives strong new tidally-induced substructure in the two highly rotationally supported models. A strong bar and bold, grand-design, two-armed spiral forms in the `Hi-Rot' model. Even the `Med-Rot' model develops a bar but it is disk-wide, and much more diffuse and fat than in the `Hi-Rot' model. Once again, the `Lo-Rot' model remains smooth and featureless. 

\vspace{2mm}
\noindent {\it{Near second apocenter (t=3.2-3.9~Gyr):}}
Most of the models show little evolution in their appearance since shortly after the second pericenter. However, the one exception is the `Hi-Rot' model. During this time period, the stars that were in the dense spiral arms at t=3.2~Gyr are stirred by the short, dense bar near the disk center. As they interact with the non-axisymmetric bar potential, they often form short-lived over-densities at the ends or on either side of the bar (the so-called Lagrangian points), and sometimes are visible at all four points simultaneously. We also note the presence of a short-lived ring-like structure, surrounding the ends of the bar, in orbit S3 of the `Hi-Rot' model (not shown in this figure).

\vspace{2mm}
\noindent {\it{Final appearance (t=6.8~Gyr):}} 
In general, the models do not evolve significantly since the previous snapshot at t=3.9~Gyr. There are two new pericenter passages during this time but they are not as deeply plunging as the first two pericenter passages, and the tidal shocking is quite weak in comparison. However, the `Hi-Rot' model shows reduced substructure, in part due to the many gigayears since the second pericenter passage, but also because the disk is now thicker and hotter which provides additional stability against forming new tidally-induced structures. The heating up of the disk with time is partly physical (due to tidal stirring) but also partly numerical (due to particle scattering). In the absence of numerical heating, later pericenter passages could potentially continue to induce spiral features. Although, we speculate that this is unlikely to change our results significantly as in general tidal stirring is found to be much more dominant than numerical heating in thickening their stellar disks. In any case, this further underlines the need for a stellar disk to be highly rotationally supported to generate tidally induced spiral-like features.

We measure the evolving rotation curve of the `Hi-Rot' model. Initially it peaks at $\sim$90~km s$^{-1}$ at a radius of $\sim$5~kpc (see left panel of Figure \ref{ICs_diskdyn}). In response to harassment, the peak rotation velocity falls to $\sim$80~km s$^{-1}$ at R$\sim$4~kpc after 80\% of the dark matter has been stripped, and $\sim$50-60~km s$^{-1}$ at R$\sim$2~kpc after 90\% of the dark matter has been stripped. We get similar results for all our strong harassment orbits (Orbit S1-S4). 
This general behaviour (i.e., reducing peak velocity with peak radius becoming closer to the center as harassment continues) was also reported for the circular velocity curve of harassed model dwarf galaxies in \cite{Smith2015} (see their Fig. 8). As explained in that study, this occurred as a result of the reducing density of dark matter surrounding the stellar disk. However, here we consider the rotation curve rather than the circular velocity, thus a combination of reduced dark matter density and increasing dispersion support must be acting in tandem to alter the rotation curve.

\vspace{2mm}
Qualitatively speaking, we note a striking similarity in the types of tidally induced features visible in our `Hi-Rot' model disk and those observed in the residual maps of our observed sample (e.g., see Figure \ref{residuals_sample}). This is not unique to `Orbit S1'. Similar features appear in the `Hi-Rot' model disk following pericenter passages for all our strong harassment orbits (Orbit S1-S4). And yet, we emphasise that the `Hi-Rot' model has a much higher degree of rotational support (V$_{\rm{peak}}/\sigma_0=2.8$ and 2.7, respectively) than that observed in real cluster dwarfs, where typically V$_{\rm{peak}}/\sigma_0=0.05$-1.0 \citep{vanZee2004, Toloba2009, Rys2014}. We propose a possible explanation -- perhaps only a small fraction of their disks is present in the form of a highly rotationally supported disk. Thus, measurements of the {\it{total}} stellar dynamics are only weakly influenced by the presence of this highly rotationally supported disk, as it makes up only a small fraction of the total light. Indeed, we note that only a small fraction of the total disk light (2.2-6.4\%) is found to present spiral-like features, with the vast majority being in a smooth and diffuse component. We extend on this idea and discuss the implications in the Discussion section (Section \ref{sec:discussion}).

\subsubsection{Induced Features Require Strong Harassment}
The results we have presented so far focus on the response of our model galaxies to orbits resulting in unusually strong harassment (Orbit S1-S4). We now consider the effects of more typical orbits -- Orbit W1, W2, W3 and W4. As we shall show, the features that are induced in the disks of the model galaxies are much weaker than for the strong harassment orbits. Therefore, we only model the `Hi-Rot' disk model response to these weaker orbits, as this is the model that is most sensitive to tidally triggered features (e.g., see left column of Figure \ref{diskresponsefig}). And, as we will see, even in this model we observe little morphological response to the `weak harassment' orbits.

\begin{figure}
\includegraphics[width=\columnwidth]{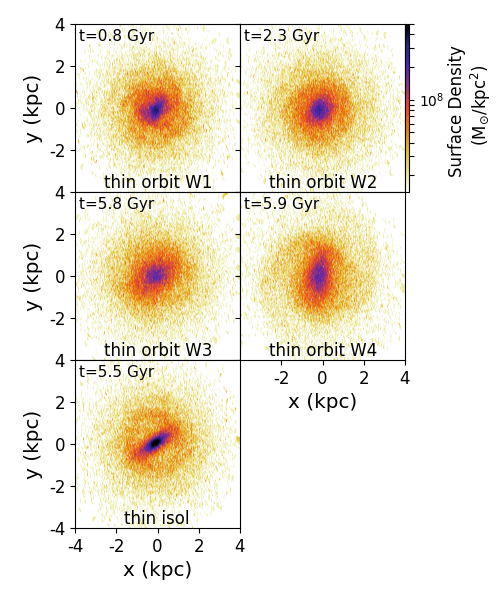}
\caption{Response of the `Hi-Rot' model to weak harassment orbits. All models are shown at 200-300~Myr after the latest pericenter passage.}
\label{diskresponseweakfig}
\end{figure}

In Figure \ref{diskresponseweakfig}, we show surface density maps of the stellar disks of the `Hi-Rot' model to the four weaker harassment orbits. We show each disk within 200-300~Myr after the latest pericenter passage, in order to see them when the tidally induced features are at their maximum. In the case of Orbit W1, W2, and W3, there is very little indication of any significant tidally triggered features. Orbit W4 does show indications for a weak spiral structure, although it is very mild in comparison to the induced spirals seen for the strong harassment orbits (e.g., see $t=1.6$~Gyr snapshot of the `Hi-Rot' model in Figure 
\ref{diskresponsefig}).

We also compare our harassed models to that of the `Hi-Rot' model evolved in isolation for the same duration. Curiously, the isolated model has undergone a disk stability and formed its own bar, even in the absence of tides, although it is quite a minor bar and there is little evidence for spiral structure. A similar result is also reported in \cite{Kwak2017}. We only see this type of non-tidally induced bar-instability form in the `Hi-Rot' model, which suggests that it requires a very high degree of rotational support to occur, much higher than that typically observed in cluster early-type dwarfs. Nevertheless, it is interesting to note that the disks on the weaker harassment orbits do not form this type of bar. This suggests that, even if the tidal disturbances are not sufficient to form spiral features, they can gently stir up the stars and disrupt the orbits of stars that would otherwise form the bar instability.

In summary, the weaker harassment orbit tests demonstrate that to generate tidally induced structure, such as clear spiral arms, requires both high rotational support within their disks combined with strong harassment.

\section{Discussion}
\label{sec:discussion}
We have conducted a study of how the effects of harassment depend on stellar disk dynamics. We find that stellar mass loss and stellar disk thickening do not show a strong dependency on disk dynamics. In contrast, we find that the tidally induced features are highly sensitive to the amount of rotational support. In particular, a high rotationally supported stellar disk like that of our `Hi-Rot' model produces spectacular tidally-induced features such as spirals, rings and bars in response to the tidal shocks occurring at pericenter passages of the cluster core. Qualitatively speaking, these features bear a striking resemblance to the features we detect buried within the dominating light of a diffuse disk in our cluster early-type dwarfs (see Section \ref{sec:residmeas}).

\subsection{Do Some Early-Type Dwarfs Have a Small Fraction of their Stellar Disks in a Thin Cold Component?}

One plausible explanation is that some early-type dwarfs, such as those with detected features, might contain a highly rotationally supported (i.e., cold) stellar disk hidden within a more massive and dispersion supported (i.e., hot) stellar disk. In this scenario, tidal triggering from a cluster pericenter passage could induce tidal features in the cold component only, without generating features in the hotter component. For example, the `Lo-Rot' model in Figure \ref{diskresponsefig} remains featureless despite undergoing the same tidal shocking which induces clear features in the `Hi-Rot' model. If the `Hi-Rot' component is only a small fraction of the total mass of the disk, then observations of the stellar dynamics would be dominated by the light from the brighter and more massive thick hot component of the disk, which could explain why the rotation support is observed to be typically much lower in cluster early-type dwarfs compared to our `Hi-Rot' model. The dominance of the light from the thick component could also explain why most early-type dwarf galaxies are observed to have quite thick disks \citep{Lisker2006b,SanchezJanssen2016}. 

To test this idea, we created one additional model galaxy called the `Mixed model', where the stellar disk was split into two components; 20\% of the disk mass is in a `Hi-Rot' component (`Mixed Hi-Rot'), and 80\% is in a `Lo-Rot' component (`Mixed Lo-Rot'). The `Mixed' model is then subjected to strong harassment on Orbit S3.

\begin{figure}
\includegraphics[width=\columnwidth]{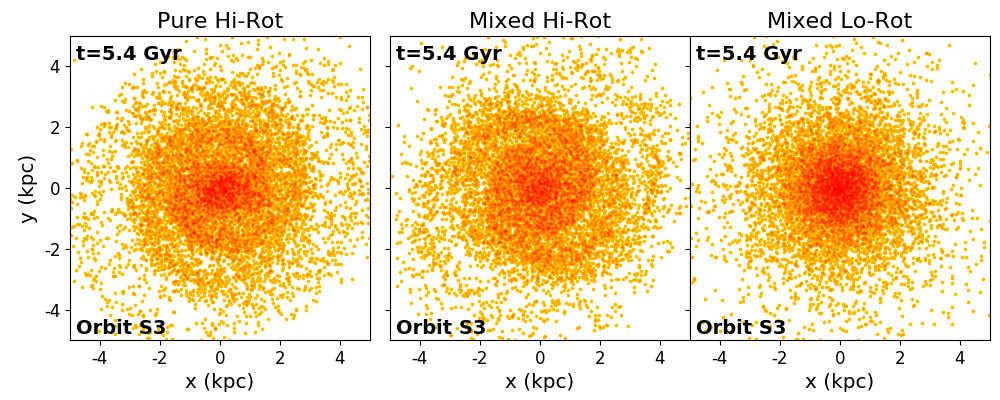}
\caption{Response of a combined `Hi-Rot' and 'Lo-Rot' disk model (the so called `Mixed model') to harassment, compared to a pure `Hi-Rot' disk model, shown for the same orbit and at the same instant. The panels are particle plots and we randomly sample the star particles to show equal numbers in each panel for a fair comparison. Tidal shocking can effectively trigger spiral features in the `Hi-Rot' component only of the `Mixed model'.}
\label{diskresponsemixfig}
\end{figure}

In Figure \ref{diskresponsemixfig}, we plot the face-on star particle distributions for the pure `Hi-Rot model' and compare it to the response of the two components of the `Mixed model' at the same instant ($t=5.4$~Gyr). As predicted, the `Mixed Lo-Rot' component remains featureless and smooth, while the `Mixed Hi-Rot' component has tidally induced spiral features. We note that the spiral features are similar but not identical to those seen in the `Hi-Rot' model, perhaps due to differing stellar densities on the disk plane.

We note that the concept of the `Mixed model' is somewhat reminiscent of the thick and thin disks observed in our own Galaxy and other nearby spiral galaxies. In our Galaxy, the thin disk is easily the most dominant component of the two components. But, in early-type cluster dwarfs this could be reversed so as the thick component is the dominating component over the thin component.

\begin{figure*}
\includegraphics[width=\textwidth]{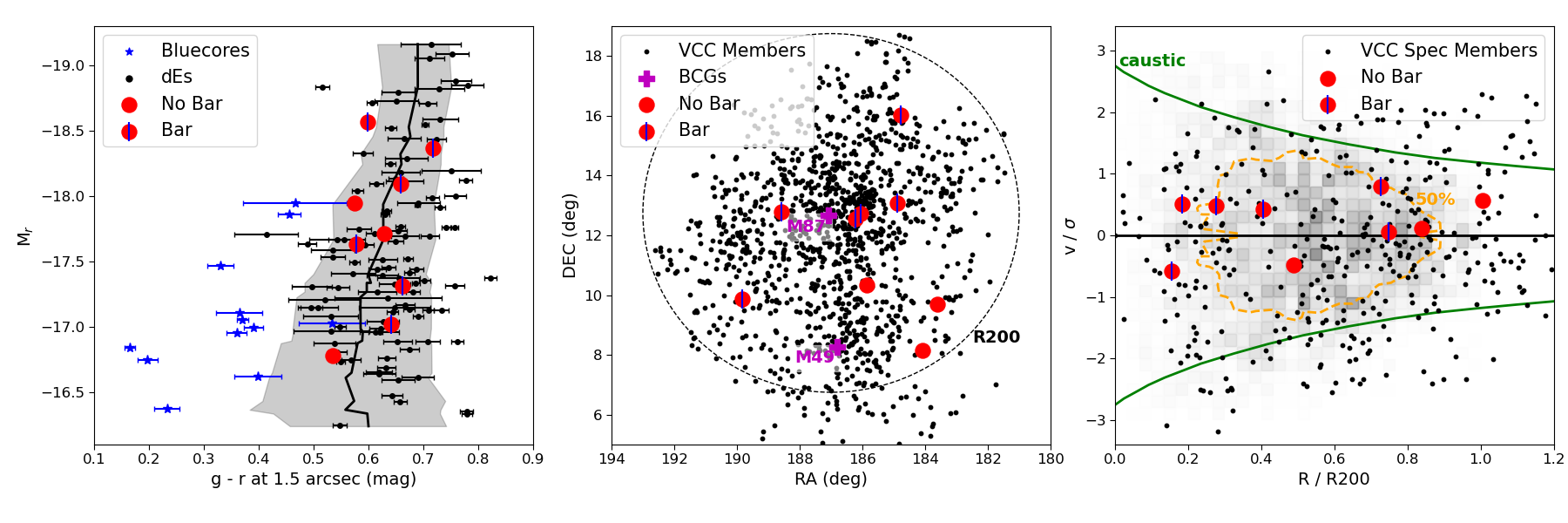}
\caption{Left panel: An optical color-magnitude diagram of Virgo cluster dwarf galaxies. The black symbols are typical dwarf ETGs and the gray shading shows the spread in the red sequence that they inhabit. Our sample is shown by the red symbols and sits within the red sequence. Blue symbols are a sample of blue-cored dwarfs from \citet{Urich2017} which clearly deviate from the red sequence. Middle panel: Location of our sample on the sky with respect to all Virgo cluster members from the Virgo Cluster Catalog (VCC; \citealp{VCC}) and the two Brightest Cluster Galaxies (BCGs; marked with purple stars). Right panel: Location of our sample in a phase-space diagram of line-of-sight velocity with respect to the cluster vs. clustocentric radius. Black points are all VCC members with radial velocity measurements. The y-axis is normalized by the cluster velocity dispersion, while the x-axis is normalized by the cluster outer radius, R200 (=1.55 Mpc; \citealp{McLaughlin1999}). Gray-scale pixels indicate the relative probability of finding galaxies that have passed cluster pericenter at least once, derived from cosmological simulations (see text for details). The orange contour encompasses half of the cosmological satellite sample. Our sample's location is more consistent with objects that have passed pericenter than objects on first infall into the cluster.}
\label{obscolfig}
\end{figure*}

\subsection{What is the Origin of the Cold and Hot Disk Components in Early-Type Dwarfs?}

The obvious manner to build a cold stellar disk component is through the formation of its stars from a disk of rotationally supported gas. For the more massive dwarfs (V$_{circ}>50$~km s$^{-1}$) considered in this study, rotational support easily dominates over dispersion support ($\sim10$~km s$^{-1}$) in their neutral hydrogen disks. Thus stars formed from this gas can inherit a similar high level of rotational support.

In order to see if the formation of the cold stellar disk component is still ongoing, we study the optical colors of our early-type dwarf sample, to try and see if there is any indication that those with spiral-features are more blue, or show different color gradients. As our deep data was observed with a white filter (to maximise image depth), we are not able to study optical colors using our own data. Therefore, here we rely on the more shallow SDSS imaging. We cross-match our sample with the early-type and blue-centered dwarf sample of \citet{Urich2017}.

Figure \ref{obscolfig} shows the results for the color-magnitude diagram (left panel). Colors are inner disk colors, and measured at 1.5 arcseconds from their centers. The early-type dwarfs are shown with black symbols, and these define a clear red sequence (black line, and gray shading). The blue-cored objects (blue stars) which are the focus of the \citet{Urich2017} study  are clearly offset from the red-sequence to bluer colors. But our sample of early-type dwarfs with spiral features (red filled circles) are fully consistent with being members of the red sequence. Similarly, the color gradients of our sample are consistent with those found in the early-type dwarfs, while the blue-cored sample deviates significantly. 

However, we note the caveat that the shallow depths of the SDSS images make it difficult to rule out that the spiral features are not slightly blue, especially as the fractions of light in the thin disk are quite small. But, given the data available to us, our sample appears fully consistent with being optically red-sequence objects, meaning there has been little evidence of significant star formation in the last one or two gigayears.

\begin{figure}
\includegraphics[width=\columnwidth]{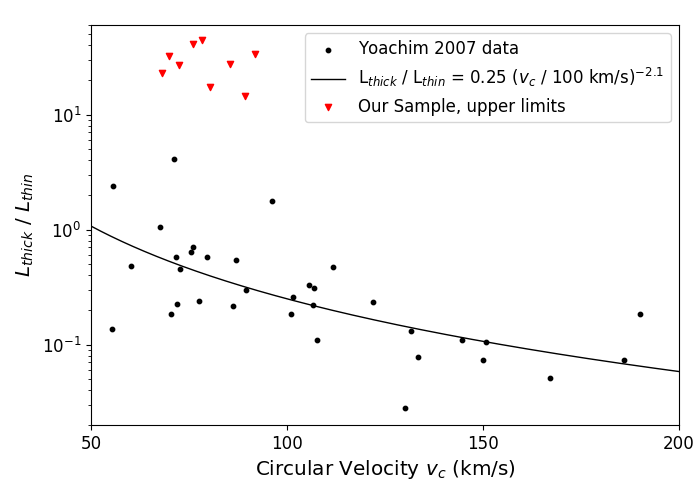}
\caption{Luminosity ratio of thick-to-thin disk measured by \cite{Yoachim2006} for a sample of near edge-on late-type spirals plotted against their circular velocities. Red data points are for our sample, but where the luminosity of the thick-to-thin disk is approximated by the light in the diffuse component divided by the light in the residual maps, measured within two effective radii. As such, the measurements for our sample should be considered upper limits (see text for details).}
\label{Yoachimfig}
\end{figure}

Another clue on the origins of the spiral features may come from the fact that our sample's residual light fractions are so low. At some point in the past, our sample of dwarf galaxies likely fell into the cluster as it hierarchically grew. Thus, it is interesting to compare their properties with those of field dwarfs. In \cite{Yoachim2006}, the fraction of thick to thin disk was measured for a sample of near edge-on field spirals, including late-type dwarf galaxies of comparable mass to our sample. The results are presented in Figure \ref{Yoachimfig}. In star-forming field galaxies (black points), there is a general trend for their thick disks to become increasingly dominant as the circular velocity reduces. For objects in a similar mass range as our sample, typically the thick to thin luminosity ratio is $\sim 0.2$-1.0.

We cannot measure the thick-to-thin luminosity ratios for our sample in the same way as done by \cite{Yoachim2006}, as these are far from being edge-on. Instead, for our sample we use the light in the diffuse component divided by the light in the residual map. We note that these are likely upper limits on the true thick-to-thin luminosity ratio (because not all of the cold disk component may present itself in the residuals). But in any case, the resulting values are substantially more thick disk dominated ($\sim 20$-40) than found in field late-type dwarfs ($\sim 0.2$-1.0) of a similar mass. As our sample is near face-on, we use a typical V$_{\rm{c}}$ value for early type dwarfs of their absolute magnitude\footnote{In practice, we make a linear fit to the observed trend in \cite{Rys2014} between V$_{\rm{c}}$ and M$_{\rm{R}}$ (V$_{\rm{c}}$=14.4$\times$M$_{\rm{R}}$+327.53).}.

Although \cite{Yoachim2006} find the thick-to-thin luminosity ratios are a function of galaxy mass, we do not find a clear trend with galaxy luminosity in Figure \ref{residualresultsfig}. However, the small range of luminosity probed by our sample, combined with the ability for an individual galaxy to change the amount of light in its residual with time in response to tidal triggering, could make detecting such a trend very challenging without a significantly larger sample.

\vskip 0.15cm
\noindent {\bf{Origin Scenarios:}}  To try to understand why our cluster dwarfs appear to have significantly higher thick-to-thin disk luminosity ratios than field dwarfs of comparable mass (see Figure \ref{Yoachimfig}), we consider three possible origin scenarios, although with our limited sample size we cannot easily distinguish among them.

\vskip 0.1cm
\noindent {\it{Nature:}}
Our sample may have recently fallen into the cluster, and had their star formation quenched by the cluster environment at least a few gigayears ago. This would leave behind the observed optically red stellar disk. But, without some form of morphological transformation, the stellar disk could be quite cold, like in their field counterparts. Given that the thick-to-thin disk fraction of field dwarfs of similar mass is so much lower than in our sample, they would have to suffer a fairly rapid morphological transformation as they are quenched, which converts the majority of their thin disk into a thick disk. Harassment could play a role in this process (e.g., Figure \ref{diskthickfig}). Or, if the dwarf galaxies were previously gas rich, the removal of their gas via ram pressure could also contribute to disk thickening \citep{Smith2012}. The need for a rapid morphological transformation could be somewhat alleviated if dwarf galaxies were first preprocessed in the galaxy group environment prior to entering the cluster \citep{Joshi2017,Han2018,Bidaran2020}.

\vskip 0.1cm
\noindent {\it{Nurture:}}
The relative fractions of the thick and thin disks could have been established prior to cluster infall by similar processes as proposed in giant galaxies. For example, minor mergers are thought to play a central role. Due to their collisional nature, the stars formed in pre-merger building blocks will inevitably tend to build a thick disk or stellar halo, while gas will tend to settle into a thin disk. Thus, to produce the very high thick disk fractions we observe, it firstly requires the dwarfs to undergo mergers with even lower mass galaxies. Indeed, there is a growing body of evidence for dwarf-dwarf mergers in low density environments \citep{Stierwalt2015,Paudel2018} and in clusters \citep{Paudel2017}. According to cosmological simulations \citep{RodriguezGomez2015}, dwarf galaxies of this mass are typically expected to suffer at least one merger of mass ratio greater than 1:10 since redshift $z=3$. Secondly, it requires the building blocks of our galaxy sample to be predominantly gas poor as this brings more stars and less dissipative gas that could potentially rebuild the thin disk. Similarly, there must be limited external gas inflow from the cosmic web \citep{Chun2020}.

\vskip 0.1cm
\noindent {\it{Progenitor bias:}}
Given that cluster early-type dwarfs likely entered the cluster a long time ago, it may be unfair to compare them to today's star-forming field dwarfs. Perhaps at the time of their infall, field dwarfs had insufficient time to grow their thin disks, and those that infall into clusters were quenched, thus preserving their disks in this manner. Meanwhile, those that remain in the field, continued to grow their thin disks through star formation, causing a growing offset between the thick-to-thin disk fractions of the field dwarfs with respect to those that fell in.

\subsection{How Many Cluster Early-Type Dwarfs Host a Cold Disk Component?}
In \cite{Lisker2006}, it is estimated that the fraction of early-type dwarfs in Virgo with disk-like features goes from $\sim50$\% at the bright end (M$_{\rm{B}}=-17$) to less than 5\% at the faint end (M$_{\rm{B}}>-15$), although not all the dwarfs with disk features present spiral-arm features. Thus, the presence of disk-like features is very common in bright early-type dwarfs. We note that there is a similar trend for increasing dominance of the thin disk component with increasing mass in late-type galaxies as well (see Figure \ref{Yoachimfig}). However, the fact that the SDSS imaging is relatively shallow compared to the observations in this study means it is unclear how much of the decrease in fractions at the low luminosity end is due to observational limitations and how much is physical.

Our simulations show that tidal triggering can reveal formerly smooth thin disks. So what causes the tidal triggering? In principle, there could be several possible sources of tidal triggering, including mergers with lower mass companions, and high speed encounters with other cluster members. But, given the fact that our sample are all spectroscopically confirmed cluster members, the cluster potential itself is an obvious and likely candidate for causing tidal triggering. Our sample's location in a clustocentric velocity vs radius phase-space diagram also supports this concept (see right panel of Figure \ref{obscolfig}). Our observed sample (red symbols) is found within one-sigma of the cluster mean velocity, which is inconsistent with a population of galaxies that are on first infall \citep{Oman2016, Rhee2017}. Using a pre-existing cosmological simulation\footnote{A simulation of a single 120~Mpc h$^{-1}$ cosmological box was conducted with Gadget-3 \citep{Springel2005}, using cosmological parameters: $\Omega_m = 0.3, \Omega_{\Lambda} = 0.7, \Omega_b = 0.047$ and $h_0 = 0.684$. We identify 42 cluster mass halos.},  we identify the typical location in phase-space where halos that have passed cluster pericentre at least once are found. The gray-scale pixels indicate their relative probability, and the orange contour encloses 50\% of the total sample. The location of our observed galaxies is generally consistent with having passed pericenter at least once.

In our simulations, we note that it is pericenter passages which drive the formation of the spiral- or bar-features and only a single pericenter passage is required. The tidal shocking itself could arise from a combination of the cluster potential and high speed encounters with other cluster members (which also occur more commonly near the cluster center). 

We also split our sample based on whether they have bars in their residual maps. Red symbols with a blue vertical bar in Figure \ref{obscolfig} indicate those classified as having bars. Their individual morphological classifications can be found in Table \ref{obssample_tab}, and the classification is done using a Fourier mode analysis (as described in detail in our companion paper, Brought to Light I: Michea et al. 2021). There is tentative evidence that the barred galaxies may slightly tend to be found closer to the cluster center although, given the small number statistics of our sample, we cannot conclude this more firmly.

In terms of the possible numbers of cluster early-type dwarfs with a cold stellar disk component, we note that some early-type dwarfs may have not reached pericenter yet. Others may not have suffered sufficient tidal shocking at pericentre. For example, our `weak harassment' simulations (pericenter distances $\sim$500~kpc) were not strongly triggered. In other words, the fractions of early-type dwarfs with disk-like features reported in \cite{Lisker2006} may represent lower limits on the true number of objects containing a cold disk component. Given that up to half of the more luminous Virgo early-type dwarfs (M$_{\rm{B}}=-17$) are reported to show visible disk features in \cite{Lisker2006b}, the true number of objects containing a cold disk component (whether they display visible features or not) may be very high for bright dwarf galaxies.

\subsubsection{How Else Might We Detect a Cold Disk Component with Early-Type Dwarfs?}

If many early-type dwarfs with cold disk components remain hidden, we ask the question: what alternative means are available to detect such objects in the absence of tidal triggering? As the thin disk component is dominated by rotation support, in principle it may be possible to detect a separate cold component from the stellar dynamics of the disks (i.e., with slit-spectra or IFU data). However, ideally the disk would have to be close to edge-on, to maximize the difference that would be observed between the two components. But early-type dwarfs with known spiral arm features tend to be closer to face-on, as this makes it easier to detect the features. Thus, it might require a blind and expensive survey of a large sample of edge-on early-type dwarfs. An easier alternative might be to search for near edge-on early-type dwarfs in well-resolved imaging, and then attempt to fit their disk profiles vertically out of the disk plane. An edge-on view would also boost the surface-brightness of the thin component.

\section{Summary \& Conclusions}
In our companion study (Brought to Light I: Michea at al 2021), we present deep optical imaging of nine early-type dwarfs in the Virgo cluster. Using a novel approach that combines several existing techniques, we split their light into a diffuse component, and a map containing the residual features. We find the residual maps contain spectacular spiral galaxy-like features including bold spiral arms, or tightly wound fine spiral structure, and bars. These features are partially obscured underneath a smooth, diffuse component that dominates the light. We find that these spiral-features typically contribute only 2.2-6.4\% of the total light, when measured within two effective radii of the galaxies. In this paper, we combine our observations with high resolution numerical simulations of harassment by a cluster potential to try to better understand the origins of these spiral galaxy-like features. 

We find that, qualitatively speaking, our models can reproduce the observed features in a scenario where close pericenter passages of the cluster core drives tidal shocking of their disks. However, it is a crucial requirement of this scenario that a cold, high-rotation support stellar disk be present from which the features form when they are tidally shocked. The amount of rotational support required (V$_{\rm{peak}}/\sigma_0 \sim2.5$-3) is much higher than is typically observed spectroscopically in cluster early-type dwarfs \citep[V$_{\rm{peak}}/\sigma_0=0.05$-1.0;][]{vanZee2004, Toloba2009, Rys2014}. 

To resolve this conundrum, we propose that some early-type dwarfs might contain just a fraction of their mass in a cold, highly rotationally supported disk. This cold component is hidden in a more massive hot, low rotationally supported disk that dominates the disk mass. Our simulations demonstrate that, in this scenario, tidal triggering by the cluster pericenter passage can effectively reveal the presence of the cold component by generating spiral-galaxy like features in it while simultaneously leaving the dominating thick, hot component smooth and featureless.

We cannot be certain what the origin of this cold stellar disk component is. However, the idea that dwarf galaxy disks might contain a thick and thin stellar disk component has already been demonstrated in the edge-on imaging of late-type disks by \citet{Yoachim2006}. In their sample, star-forming field dwarfs of similar luminosity to our sample are shown to contain a thick and thin disk in roughly equal proportions. As our sample is closer to face-on, we cannot measure the thick-to-thin disk ratios in the same way. But with residual fractions of only a few percent, our sample appears to be much more dominated by a diffuse component than in the star-forming field dwarfs. This could arise because the star-forming field progenitors of our sample were much more thick disk dominated at the time of infall into the cluster. Or, perhaps, field dwarfs rapidly converted their cold stellar disks into a hotter component as they were quenched in the cluster environment. Alternatively, the merger building blocks of cluster dwarfs could have been more gas poor, perhaps due to the denser environment in which they are located, resulting in a more thick disk dominated disk from the outset (see the Discussion section for more details).

Measurements of the optical colors and optical color-gradients of our observed sample demonstrate they are consistent with other red-sequence dwarf galaxies, that do not present spiral features. Thus, star formation probably ceased in their cold disk component at least one or two gigayears ago\footnote{With the caveat that the optical colors are based on SDSS images which are fairly shallow compared to our own imaging.}. Despite the shallow imaging, \cite{Lisker2006b} finds as many as half of their bright early-type dwarfs (M$_{\rm{B}}=-$17) contain disk-like features. And these may only be the tip of the iceberg.
Our simulations show that triggering of spiral-features requires at least one deeply plunging passage of the cluster core (R$_{\rm{peri}}<0.25$~R$_{\rm{vir}}$) for tidal shocking to be sufficiently strong. Thus, many early-type dwarfs may not have suffered sufficient tidal shocking for their cold disk component to be revealed, meaning such a component may be very common (even it has yet to be unveiled) in luminous early-type dwarfs.

To better test this hypothesis, it is clear that we require a much more complete set of detections of cold disk components in early-type dwarfs, in order to better constrain their properties and distinguish between the origin scenarios. While tidal triggering is one manner in which we might detect such a cold component, we propose that other detection methods may exist. Under the assumption that there are two distinct components (i.e., a thin and thick disk), it may be possible to detect the thin component directly in near edge-on early type dwarf, in the form of a double vertical profile. The stellar dynamics of near edge-on dwarfs may also reveal two separate components, if present. However, if there are not two distinct components, but rather more a continuous spectrum moving from cold stars to hot stars, then determining this using these alternative detection methods may be more challenging.

In the future, deep multi-wavelength imaging or spectroscopy of early-type dwarfs with hidden spiral features may help to determine the time since star formation last occurred in their spiral arms, which in turn will help constrain our understanding of the cold disk's origin. Cosmological simulations could also help to constrain the origin of this component. 
Nonetheless, we note the need for very high spatial resolution to resolve the thin, cold disks vertically, combined with a dense enough environment to cause the tidal triggering, which is a highly numerically challenging combination. We hope that with deep imaging of a large sample of cluster or group dwarfs (e.g., in Virgo, Fornax, etc.), a larger population of early-type dwarfs with hidden spiral features may be discovered, and this will allow us to better constrain the properties of their host galaxies, leading to a deeper understanding of their origins.

\section*{Acknowledgements}
P.C-C. was supported by CONICYT (Chile) through Programa Nacional de Becas de Doctorado 2014 folio 21140882. J.M. acknowledges support by the Deutscher Akademischer Austauschdienst (DAAD), Universit\"at Heidelberg, and the Max-Planck-Institut f\"ur Astronomie. J.M., A.P., R.P., R.S. and E.K.G. acknowledge financial support from the European Union's Horizon 2020 research and innovation program under the Marie Sklodowska-Curie grant agreement no. 721463 to the SUNDIAL ITN network. S.P. acknowledges support from the New Researcher Program (Shinjin grant No. 2019R1C1C1009600) through the National Research Foundation of Korea.
\noindent

\bibliographystyle{aasjournal}
\bibliography{bibfile}














\end{document}